# Density-driven scattering and valley splitting in undoped Si/SiGe two-dimensional electron system

Lucky Donald Lyngdoh Kynshi[1], Umang Soni[1], Chithra H Sharma[2,3], Yu Cheng[4], Kristian Deneke[3], Robert Zierold[3], Shengqiang Zhou[4], Robert H Blick[3], Anil Shaji[1], and Madhu Thalakulam[1*]

[1] Department of Physics, Indian Institute of Science Education and Research Thiruvananthapuram, Kerala 695551, India

[2] Institut für Experimentelle und Angewandte Physik, Christian-Albrechts-Universität zu Kiel, 24098 Kiel, Germany

[3] Center for Hybrid Nanostructures, Universität Hamburg, Luruper Chaussee 149, 22761 Hamburg, 22761 Germany

[4] Helmholtz-Zentrum Dresden-Rossendorf, Institute of Ion Beam Physics and Materials Research, Bautzner Landstraße 400, 01328 Dresden, Germany

## ABSTRACT

Undoped Si/SiGe two-dimensional electron gas (2DEG) provide an ideal platform for hosting quantum-dot spin-qubits owing enhanced spin dephasing times and compatibility with standard CMOS technology. The strained Si quantum well reduces the valley degeneracy into two closely spaced ones. The existence of a near-degenerate valley state act as a leakage channel and compromises gate fidelity. A robust and uniform valley splitting across the entire chip is crucial for achieving scalability in the architecture and reliability in operation. Imperfections such as broadened interfaces, alloy disorders and atomic steps significantly compromise the valley splitting. The associated scattering mechanisms play detrimental roles in the performance of the qubits. In this manuscript, exploiting low-temperature magnetotransport measurements, we investigate the scattering mechanisms and valley splitting in a high-mobility ($\mu \approx 1.6 \times 10^6 \text{ cm}^2\text{V}^{-1}\text{s}^{-1}$) undoped Si/SiGe 2DEG. At lower carrier densities, transport is limited by remote impurity scattering, whereas at higher densities, background impurity scattering near the quantum well dominates. Both the transport and quantum lifetimes of the charge carriers increase with carrier concentration, due to the enhancement in the impurity screening. Magnetic-field-induced confinement effect also is found to improve the valley splitting. Visibility of both the spin and valley splitting is influenced by the level broadening due to electron–impurity scattering and temperature. Current-biasing measurements reveals the role of carrier heating in the visibility of valley splitting and reveal a temperature limited valley splitting of $\Delta_v \sim 100 \, \mu\text{eV}$. These results provide critical insight into scattering-dominated regimes and valley splitting in undoped Si/SiGe, advancing its potential for silicon-based quantum devices.

---

[*] madhu@iisertvm.ac.in

# I. INTRODUCTION

Spin-states of isolated electrons in gated quantum dots are one of the leading hosts to realize quantum computers [1–5]. Ever since the original proposal [6], two-dimensional electron gas (2DEG) formed either on GaAs/AlGaAs or Si/SiGe quantum well heterostructures have been the major focus [7–11]. The shorter decoherence times, owing to the Fermionic nuclei and the resulting hyperfine interaction in GaAs based material systems propelled the development of silicon-based gated quantum dots for the development of spin-qubits [12,13] . In this regard, isotopically purified $^{28}$Si/SiGe 2DEG owing to its negligible nuclear spin density have shown to eclipse spin coherence times in the range of ~1s [14] , many orders higher than that of natural silicon systems [10,15]. Early work on modulation-doped Si/SiGe heterostructures successfully demonstrated many key milestones, including the realization of gated double quantum dots [16], the investigation of spin blockade [17], tunable tunnel coupling [18], and excited-state spectroscopy [19]. However, the ionized impurities in the dopant layer of the modulation-doped structure introduce charge noise [20,21], leakage current, and cause device instability [20], thereby limiting the progress to probe intrinsic quantum phenomena [22]. In contrast, undoped Si/SiGe, free from ionized dopants, has shown improved charge stability, higher mobilities, and reduced potential fluctuations [23,24] and has been one of the main platforms for the development of electron spin-based quantum processors on silicon[ [4,25–27].

While one-and two-qubit operations have been demonstrated on these platforms [4,28–31], further technological advancements depends on the scalability of spin-qubits on these platforms which require an in-depth understanding of the intrinsic material properties, such as mobility, impurity scattering, density homogeneity, and the valley splitting energy [32]. Residual disorder due to the interface traps and background impurities can degrade mobility and contribute to spin dephasing [33]. Additionally, it is important to enhance the valley splitting arising from the strained silicon quantum well to reduce inter-valley excitations [34], to prevent spin-valley entanglement [35], and to strengthen the Pauli spin blockade [36]. Valley splitting is highly sensitive to the Si/SiGe interface roughness [37] and its enhancement relies on the impurity screening and the degree of confinement [38] arising from externally applied electric and magnetic fields [39].

While the role of scattering mechanisms on various performance matrices have been investigated in GaAs/AlGaAs systems [40,41], a comprehensive study in this regard on undoped Si/SiGe 2DEG is still wanting at large. Here, we present magnetotransport measurements on high mobility undoped Si/SiGe 2DEG, focusing on (i) density-driven scattering mechanisms, (ii) evolution of the Landau level broadening and carrier life-times as a function of the carrier density, and (iii) the influence of carrier density and magnetic field on valley splitting. We employ a Hall-bar device and measure the longitudinal and transverse resistances under perpendicular magnetic fields, down to 10 mK in temperature, in a cryogen-free dilution refrigerator. Analysis of Shubnikov-de Haas (SdH) oscillations and Hall resistances at low magnetic fields enables the extraction of carrier density and mobility, from which we identify the dominant long and short-range scattering mechanisms across the accessible density range. The magnetic field dependence of SdH amplitude further reveals quantum lifetimes and quantifies the evolution of impurity-induced Landau level broadening with carrier density. At higher fields, the SdH oscillations reveal spin and valley states, with a resolution governed by the relative magnitudes of Zeeman, valley splitting, disorder, and thermal broadening energies. We extract the valley splitting and characterize its dependence on the magnetic field and disorder broadening. Further, the intrinsic valley splitting on our device is extracted from the thermally activated behaviour of SdH oscillations.

## II. EXPERIMENTAL DEVICE SETUP

The Hall bar devices used in this study are realized on a commercially sourced Si/SiGe quantum well heterostructure wafer (Lawrence Semiconductor Research Laboratory). Figure 1(a) shows a schematic representation of the layer structure, which includes a 225nm $Si_{0.7}Ge_{0.3}$ relaxed buffer, strained 8 nm Si quantum well, and 50 nm $Si_{0.7}Ge_{0.3}$ barrier capped by a 2nm Si layer. The two-dimensional electron gas (2DEG) is located ~50 nm beneath the surface at the interface of the strained Si and $Si_{0.7}Ge_{0.3}$ layer. We fabricate a Hall bar device ($Width \times Length = 75 \times 375$ μm) using standard electron-beam lithography, followed by $SF_6 + O_2$ plasma etching. Highly doped n++ regions, serving as ohmic contacts, are realized by phosphorous ion implantation followed by rapid thermal annealing. A ~90 nm thick layer of aluminium oxide ($Al_2O_3$), deposited using atomic layer deposition at 150 °C, is used as the accumulation gate-dielectric. All measurements are conducted in a dilution refrigerator equipped with a superconducting magnet at a base-temperature of ~10 mK. Figure 1(b) shows the measurement configuration used for all measurements discussed in this study. We apply a positive bias to the top-gate to accumulate electrons in the quantum well. The longitudinal ($R_{xx}$) and transverse ($R_{xy}$) resistances are measured simultaneously using lock-in amplifiers. All the magnetotransport data discussed in this study are acquired with the field applied perpendicular to the plane of the 2DEG.

Figure 1(c) shows the transfer characteristics, source-drain current ($I_{sd}$) versus top-gate voltage ($V_G$), at a fixed source-drain bias ($V_{sd}$) of 1 mV. The device shows a sharp rise in $I_{sd}$ when $V_G$ is swept in the positive direction beyond a voltage of ~0.45 V, which is a signature of a clear turn-on and formation of a conducting channel in the silicon quantum well. We identify three distinct operational regimes, I, II, and III, in Fig. 1 (c). In regime I ($V_G < 0.45V$), the channel remained insulating because the lowest sub-band of the quantum well lies above the Fermi level, resulting in no charge carriers available for transport. In regime II ($0.45\,V \lesssim V_G \lesssim 0.7\,V$), the channel turns on at a threshold voltage $V_{th} \approx 0.45V$, and the current increases sharply with $V_G$. This behaviour arises because the conduction band-edge drops below the Fermi level, allowing electrons to populate the 2D sub-band of the quantum-well and thereby facilitating electron transport. In regime III ($V_G > 0.7V$), the current exhibits a weaker gate-voltage dependence, suggesting a screening of remote scattering impurities and a transition to the dominant background impurities scattering mechanism [42,43].

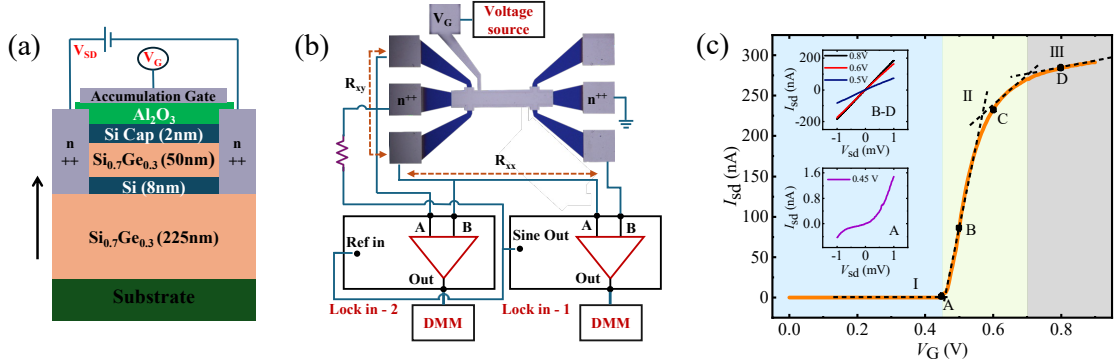

Fig. 1 (a) Schematic illustration of the undoped Si/SiGe heterostructure used in this study. The 8nm Silicon strained quantum well, 50 nm below the crystal surface is sandwiched between relaxed 225 nm $Si_{0.7}Ge_{0.3}$ layers and 50nm $Si_{0.7}Ge_{0.3}$ barrier. A highly phosphorus-doped n$^{++}$ regions, formed by ion implantation, serves as ohmic contact. A 90 nm $Al_2O_3$ layer acts as the gate dielectric between the accumulation gate ($V_G$) and the Si/SiGe wafer. (b) Schematic representation of the measurement configuration; $R_{xx}$ and $R_{xy}$ denotes the longitudinal and transverse resistances, respectively. (c) Transfer characteristics of the device. Three different operational regimes are identified: regime I ($V_G < 0.45V$ insulating), regime II ($0.45\,V \lesssim V_G \lesssim 0.7V$ electron accumulation), and regime III ($V_G > 0.7V$, background impurity dominated scattering). Bottom inset: nonlinear I–V characteristics at point A (regime I), representing the Schottky nature of the source and drain contacts. Top inset: Linear I–V curves at points B, C, and D (regimes II, and III, respectively).

We further measure the current-voltage (*I-V*) characteristics of the three regimes, as shown in the inset to Fig. 1 (c). Point 'A' (inset, bottom) shows a nonlinear *I-V* curve indicative of limited electron accumulation in the quantum well and the formation of Schottky barrier between the ohmic contact and the quantum well, whereas points B, C, and D (inset, top) show a linear *I-V* curve, confirming ohmic conduction.

## III. LOW MAGNETIC FIELD TRANSPORT MEASUREMENTS

We perform low magnetic field Hall measurements to determine the carrier concentration and mobility at various accumulation gate voltages covering all three transport regimes, B, C and D shown in Fig. 1 (c). Furthermore, we investigate the dominant scattering mechanisms from the power-law dependence of mobility on carrier concentration. Figure 2 (a) shows the low-field magnetotransport data, $R_{xx}$ and $R_{xy}$ versus magnetic field, for $V_G$ =1.4 V. We find that the $R_{xx}$ exhibit clear SdH oscillations for field values beyond ≈ 0.25T, while the

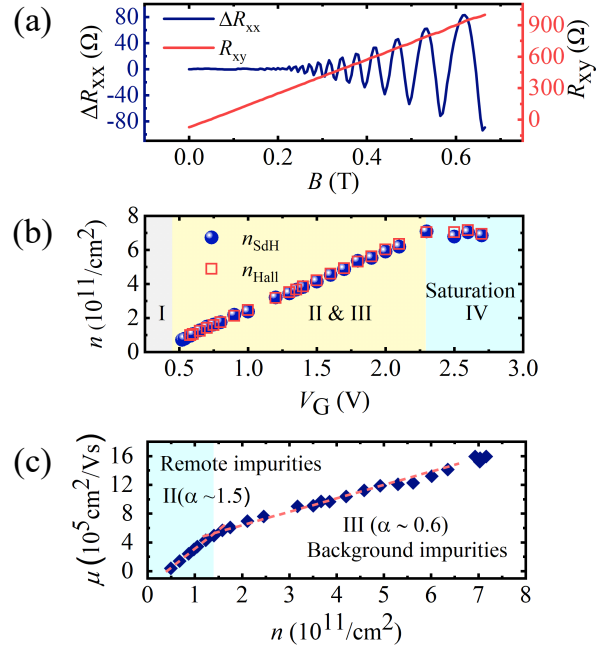

Fig. 2 (a). Shows the low-field magnetotransport measurement at $V_G$=1.4 V. SdH oscillation in $R_{xx}$ emerge at $B \approx 0.25$T, while the Hall resistance $R_{xy}$ varies linearly with the magnetic field. (b) Carrier density as a function of gate voltage extracted from Hall measurements $n_{Hall}$ (red squares) and from SdH oscillations $n_{Sdh}$ (blue circles). Regime I correspond to $V_G$ values below the turn-on gate voltage, and Regimes II, III exhibit a linear density dependence with the accumulation gate voltage ($V_G$). In Regime IV above 2.3 V (shaded blue), the carrier density saturates at $n = 7.09 \times 10^{11} \text{cm}^{-2}$. (c) Electron mobility (*µ*) as a function of carrier density (*n*) with a maximum mobility $\mu \approx 1.6 \times 10^6 \text{cm}^2\text{V}^{-1}\text{s}^{-1}$. Regimes II and III have power law dependencies of $\alpha = 1.5$ and 0.6 respectively. Remote impurity scattering dominates at low carrier concentration $n < 1.4 \times 10^{11} \text{cm}^{-2}$, while scattering due to background impurities near the quantum well dominates at higher concentration $n > 1.4 \times 10^{11} \text{cm}^{-2}$.

Hall resistance $R_{xy}$, exhibits a linear magnetic field dependence. From the $R_{xy}$, we extract the Hall carrier density $n_{Hall}$, while from the periodicity of the SdH oscillations we extract the SdH carrier density $n_{SdH}$. The extracted carrier concentrations, $n_{Hall}$ (blue circles) and $n_{SdH}$ (red squares) versus accumulation gate voltage are shown in Fig. 2 (b). The excellent agreement between $n_{Hall}$ and $n_{SdH}$, confirms the formation of a clean 2DEG and negates the possibility of any parallel conduction channel on our device. In Fig. 2 (b), we observe that in regime I ($V_G <$ 0.45V), the carrier density approaches zero, which is consistent with the absence of any measurable current in Fig. 1(c) . In regimes II and III the carrier density increases linearly up to a gate voltage of ~ 2.3 V, suggesting the effective capacitive coupling between the 2DEG and the accumulation gate [44,45]. In regime IV, beyond $V_G = 2.3$V (shaded in blue ), the carrier density saturates at $n = 7.09 \times 10^{11} \text{cm}^{-2}$ which we believe is due to the formation of a parallel channel at the Si/Al$_2$O$_3$ interface consisting of the interface states [46]. The trapped electrons at the Si/oxide interface, screens the electric field due to the accumulation gate preventing further 2DEG density modulation [43,46]. Once the accumulation gate voltage reaches this regime, blue shaded region in Fig. 2 (b), the interface traps induce hysteresis in the transfer characteristics and in the carrier density as shown in Supplementary Material SM-1.

Figure 2 (c) shows a plot of the carrier mobility against carrier concentration, extracted from the zero-field $R_{xx}$, employing Drude's model. We find that the mobility increases with carrier density owing to the enhanced screening of the Coulombic potential due to the charged impurities by the 2DEG [45,47]. The device exhibits a peak mobility of $\mu \approx 1.6 \times 10^6 \text{ cm}^2\text{V}^{-1}\text{s}^{-1}$ at a carrier density of $n = 7.09 \times 10^{11} \text{cm}^{-2}$. From the power-law dependence between the mobility and the carrier concentration, $\mu \propto n^\alpha$, we identify two distinct transport regimes, II and III, at low and high carrier concentrations, respectively. Scattering due to the remote charge impurities at the Al$_2$O$_3$/Si interface dominates in regime II ($n < 1.4 \times 10^{11} \text{cm}^{-2}$, $\mu \approx 10^5 \text{cm}^2\text{V}^{-1}\text{s}^{-1}$) with $\alpha \approx 1.5$. In regime III ($n > 1.4 \times 10^{11}\text{cm}^{-2}, \mu \approx 10^6 \text{cm}^2\text{V}^{-1}\text{s}^{-1}$), with $\alpha \approx 0.6$, the scattering due to the background impurities near the quantum well limits the mobility [48]. Following Monroe et. al [48], We estimate a background impurity density of $\sim 10^{13}/\text{cm}^3$ (see Supplementary Material SM-1) which is lower by an order in magnitude comparable to the previously reported values in a high-quality Si/SiGe heterostructure [44,49].

To investigate the nature of the remote and background scattering impurities, we analyse the transport lifetime ($\tau_t$) and quantum lifetimes ($\tau_q$), which probes the large and small angle scatterings [50,51] respectively. The transport lifetime is primarily sensitive to large-angle scattering from short-range impurities near the quantum well, and it is directly related to the experimentally measured Hall mobility in Fig. 2(c) by the relation $\mu = \frac{e\tau_t}{m^*}$, where $m^* = 0.19 m_o$ is the effective mass of electrons in silicon and $m_o$ is the mass of a free electron [52]. The transport lifetime extracted from the Hall mobility increases from 34 ps to 181 ps with increasing carrier concentration $n = 1 \times 10^{11} \text{cm}^{-2}$ to $7 \times 10^{11} \text{cm}^{-2}$ which is due to the enhanced screening of impurities by the 2DEG.

However, to account for electron-impurity scattering at all angles and to characterize the single-particle relaxation, we experimentally determine the quantum lifetime from the Landau level broadening of the SdH oscillation. Figure 3(a) shows the background-subtracted SdH oscillation $\Delta R_{xx} = R_{xx} - R_b$, versus *1/B* for $n = 1.75 \times 10^{11} \text{cm}^{-2}$ (top) to $2.45 \times 10^{11} \text{cm}^{-2}$ (middle) and to $3.5 \times 10^{11} \text{cm}^{-2}$ (bottom). $R_b$ is the background resistance (see Supplementary Material SM-2 for background subtraction). We confine our analysis to low magnetic field regime, before the onset of spin-splitting, to avoid amplitude modulation of the SdH oscillations [53]. We also did not consider to data sets with $n < 1.75 \times 10^{11} \text{cm}^{-2}$, owing to the weak visibility of the SdH oscillations. Following the Coleridge's method to extract the quantum lifetime [50,54], we fit the peak amplitude of Shubnikov-de Haas (SdH) oscillations $\Delta R_{xx}$ to

$$\Delta R_{xx} = 4R_o X(T) \exp\left(-\frac{\pi}{\omega_c \tau_q}\right) \quad (1)$$

Where $R_o$ is the zero-field resistance, $\omega_c = eB/m^*$ is the cyclotron frequency, and $X(T)$ is the thermal damping factor given by

$$X(T) = \frac{\frac{2\pi^2 k_B T}{\hbar \omega_c}}{\sinh\left(\frac{2\pi^2 k_B T}{\hbar \omega_c}\right)} \quad (2)$$

Here, $K_B$ is the Boltzmann constant, $e$ is the elementary charge, and $\hbar$ is the reduced Planck's constant. Figure 3 (b) shows a plot of $\ln\left(\frac{\Delta R_{xx}}{4R_o X(T)}\right)$ versus *1/B*. The linear dependence, following equation (1), confirms the uniform carrier density in the 2DEG at low magnetic fields [50]. The slope yields a quantum lifetime in the range of $1.688 \text{ ps} - 3.32 \text{ ps}$ which is comparable to previously reported values [44].

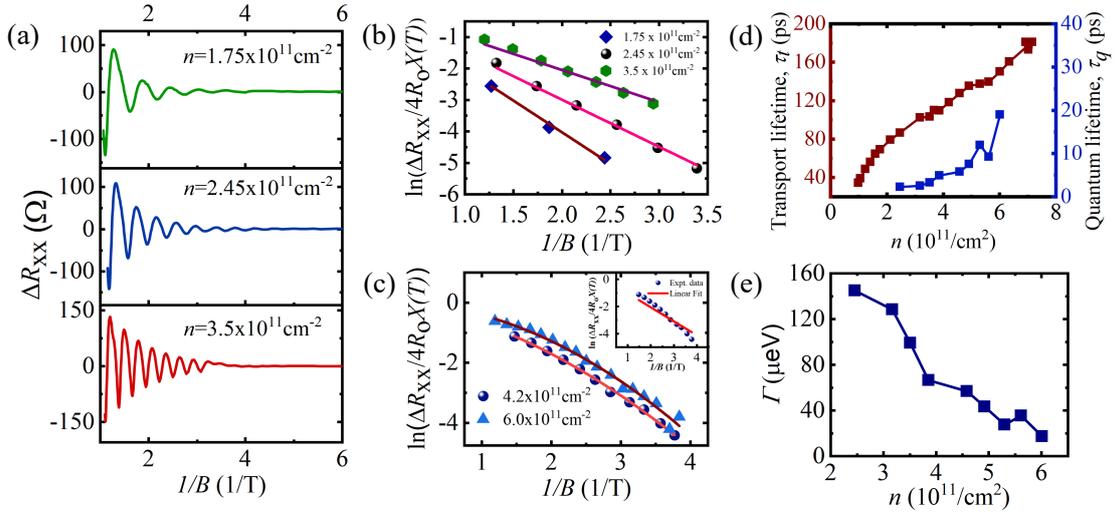

Fig. 3 (a) SdH oscillation with *1/B* after background subtraction at $n = 1.75 \times 10^{11} \text{cm}^{-2}$ to $3.5 \times 10^{11} \text{cm}^{-2}$. (b) Logarithmic plot of SdH oscillation amplitude versus *1/B*, showing experimental data (points) and corresponding linear fit (solid line). (c) For $n > 4.2 \times 10^{11} \text{cm}^{-2}$, the amplitude variation with *1/B* deviates from the linear fit, as shown in the inset. The experimental data were fitted with a modified equation accounting for the density inhomogeneity, showing good agreement with the data (solid line). (d) Transport lifetime and a quantum lifetime versus carrier concentration. (e) Landau level broadening against carrier concentration extracted from the quantum lifetime.

We conduct a similar exercise of the extraction of quantum life-time at higher carrier concentrations i.e. $n \geq 4.2 \times 10^{11} \text{cm}^{-2}$, as shown in Fig. 3 (c). We find that the experimental data deviates from the linear dependence, in contrast to that observed in Fig. 3 (b). For instance, the logarithmic plot of the SdH peak amplitude against *1/B* for $n = 4.2 \times 10^{11} \text{cm}^{-2}$ in the inset to Fig. 3 (c) shows a clear deviation from the linear dependence observed for lower carrier densities. Although one can extract the quantum lifetime from such a fit, the resulting value is significantly lower relative to the actual lifetime [51]. Previous studies have attributed this nonlinearity in high-quality (2DEGs) to spatial fluctuations in the local carrier density across the sample [50,51,55]. To account for this effect, we incorporate an additional term $\left(\frac{\pi^2 \hbar \delta n}{m^* \omega_c}\right)$ in the exponential factor of Eq.(1), where $'\delta n'$ represents the density inhomogeneity term [55]. Using $\tau_q$ and $\delta n$ as fitting parameters, we obtain an excellent agreement between the experimental data (scatter plot) and the fitted curves (solid lines), as shown in Fig. 3 (c). Applying the same fitting procedure across all carrier concentrations, we obtain a relative density variation $(\delta n/n_o)$ of ~ 1.2%, where $n_o$ is the carrier density from the Hall measurement. The extracted quantum lifetimes are plotted along the transport lifetime in Fig.

3 (d). Both the transport lifetime (34 ps $\leq \tau_t \leq$ 181 ps) and quantum lifetime (1.6 ps $\leq \tau_q \leq$ 19 ps) increase with the carrier concentration, reflecting the enhanced impurity screening by the 2DEG. The ratio $\frac{\tau_t}{\tau_q} \gg 1$ but it approaches to unity at higher carrier concentrations (see supplementary Material SM-3). This suggests that the low angle (long-range scattering due to impurities at the dielectric interface) dominates at lower carrier concentrations, whereas the large angle scattering (short-range impurities near the quantum well) dominates at higher carrier concentrations, consistent with the power law observation $\mu \propto n^\alpha$ in Section III, Fig. 2(c)

From the quantum lifetime, we estimate the scattering induced broadening of the Landau levels, using $\Gamma \approx \frac{\hbar}{2\tau_q}$, shown in Fig. 3 (e) [51,54]. The broadening of the Landau levels due to the electron-impurity scattering decreases from $\approx 144\mu eV - 17.3\mu eV$ with increasing carrier concentration, providing evidence for the effective screening of the impurities by the 2DEG.

## IV. HIGH MAGNETIC FIELD TRANSPORT MEASUREMENTS

The quantum lifetime broadening of Landau levels limit the observation of spin and valley splitting at lower magnetic fields, and we move onto high-field SdH oscillations to investigate the same on our device. Figure 4 (a) shows the $R_{xx}$ and $R_{xy}$ versus magnetic field, upto $B = 6.5$ T, for $n \approx 4.2 \times 10^{11} cm^{-2}$. The observation of clear SdH oscillations in $R_{xx}$ and concurrent quantized plateaus in $R_{xy}$ confirms the formation of well-resolved Landau levels. We note here that the plateaus corresponding to lower filling factors, $\nu = 1,2$, lie outside the accessible magnetic field range in our measurement system. The left inset to Fig. 4 (a) shows an energy-level diagram showing the evolution of the spin and valley splitting in Si/SiGe 2DEG. We find that the SdH oscillations beyond a magnetic field $B_S \approx 1T$ shows the emergence of spin-splitting. From which, we estimate an effective g-factor $g^* \sim 2.93$, higher than the bulk g-factor value. This corresponds to an enhanced spin-splitting $E_s = g^* \mu_B B \approx 159 \mu eV$ at 1 T (Details of g-factor extraction and its variation with carrier concentration are shown in supplementary Material SM-4). A similar enhancement of the g-factor has been reported previously and is attributed to the exchange interactions of electrons [44].

For magnetic fields beyond $B_V \approx 2T$, we observe the evolution of another splitting in the SdH oscillations corresponding to lifting of valley degeneracy. The splitting of SdH peaks

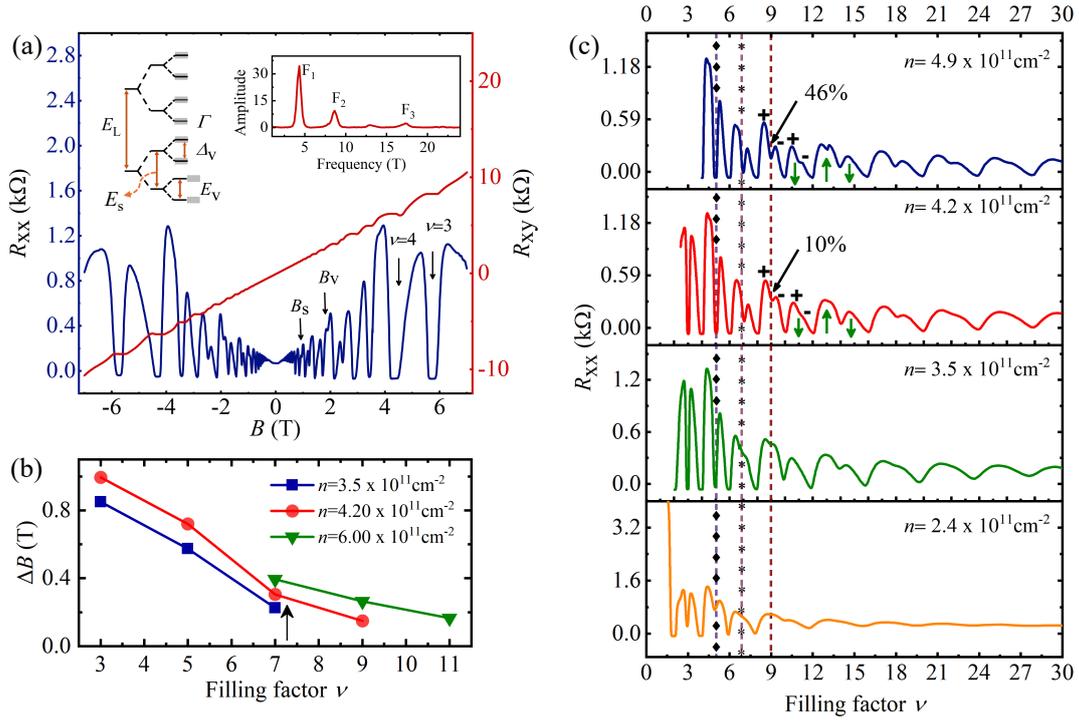

Fig. 4(a) $R_{xx}$ and $R_{xy}$ with versus magnetic field at $n \approx 4.2 \times 10^{11} cm^{-2}$. $B_s$ and $B_v$ mark the onset of spin and valley splitting, respectively. Left Inset: Schematic energy level diagram showing the Landau level splitting into spin and valley sublevels with corresponding energies $E_L$, $E_s$, and $E_v$. The effective valley splitting is $\Delta_v = E_v - \Gamma$, where $\Gamma$ denotes the Landau level broadening. Right inset: FFT spectrum showing peaks $F_1$, $F_2$, $F_3$ displaying the Landau, spin, and valley contributions. (b) Peak-to-peak spacing between adjacent oscillations at odd filling factors corresponding to valley splitting. (c) Oscillations of $R_{xx}$ with filling factor at different carrier concentrations. The splitting at $v = 9$ (dotted lines) shows an increase in relative resistance dip with increase in carrier concentrations and similar observations are observed at $v = 5$ & $7$ (represented by star and diamond markers). (↑,↓) and (+,−) in top and middle panels represent the spin and valley states, respectively

corresponding to the odd filling factor is due to the valley degree of freedom. Assuming that both the spin and valley splitting become resolvable only when their respective energy gaps exceeds the landau level broadening, $\Gamma$, we estimate a valley splitting $E_V(2T) \approx \Gamma \approx E_S(1T) \sim 159 \mu eV$. The broadening $\Gamma$, estimated from the onset of spin splitting is higher than that we extract from the quantum lifetime suggesting that, in addition to the impurity scattering, the SdH oscillations also suffer thermal broadening effects. The Fast Fourier Transform of the resistance $R_{xx}$ against $1/B$, shown in the right-inset of Fig. 4 (a), displays three distinct peaks $F_1 = 4.32$ T, $F_2 = 8.64$ T, and $F_3 = 17.29$ T. The frequency $F_1$ is the fundamental frequency corresponds to the Landau level where as the frequencies $F_2$ and $F_3$ corresponds to the spin and valley states.

Figure 4 (b) shows a plot of the peak-splitting for the valley states, $\Delta B$, against the filling factor. We find that the splitting ($\Delta B$) reduces (increases) with the filling factor (magnetic field). This enhancement in valley splitting, manifested as $\Delta B$, with the magnetic field is a result of the stronger magnetic field-induced confinement, from ~ 18 nm to ~ 10 nm, of the electron wavefunction [39].

Interface roughness and associated scattering have shown to influence valley splitting suggesting that the carrier concentration and impurity screening can play a significant role in the observation of valley splitting [37,56,57]. In this regard, in Fig. 4 (c) we investigate the role of carrier concentration on valley splitting. The current through the device is set to 100 nA, ensuring any thermal contribution to the Landau broadening remains minimal. As we increase the carrier concentration from $n = 2.4 \times 10^{11} \text{cm}^{-2}$ to $4.9 \times 10^{11} \text{cm}^{-2}$, the SdH oscillations become more pronounced by lowering the Landau level broadening as a result of the enhanced screening of the impurities by the 2DEG. The notations ($\uparrow, \downarrow$) and ($+, -$) in Fig. 4(c) top and middle panels, correspond to the spin and valley states, respectively. Focusing on the valley splitting at the filling factor $v = 9$, represented by dotted lines in Fig. 4 (c), we observe that the relative dip in the resistance increases from 0% to 46%, suggesting the enhancement of valley splitting with carrier concentration. A similar enhancement of the valley splitting has been reported and attributed to the enhanced overlapping of the electron wave function with Ge atoms at the Si/SiGe interface [58,59]. At low carrier densities, shown in Supplementary Material SM-4, we find that the valley splitting is not well resolved. We also observed a similar enhanced valley splitting at other filling factors, $v = 7$ and 5 (represented by star and diamond shape markers) in Fig. 4(c) suggesting that the resolution of the splitting depends on the Landau level broadening which in turn is influenced by the carrier concentration and screening effects in the 2DEG.

## V.     TEMPERATURE DEPENDENCE

The estimation of valley splitting energy from the onset of valley-resolved SdH oscillations is indirect, as the onset is strongly influenced by disorder-induced Landau level broadening. Consequently, the extracted splitting represents only a lower bound, with a magnitude comparable to the broadening $\Gamma$. To quantitatively extract the intrinsic valley splitting while excluding the effects of the broadening given by $\Delta_v = E_v - \Gamma$, we utilize the temperature dependence of the SdH oscillations [58]. Figure 5 (a) shows the SdH oscillations measured at a carrier density of $n = 2.98 \times 10^{11}$ cm$^{-2}$ over a temperature range between 15 mK and 850 mK. The amplitude of the oscillations decreases with temperature. Focusing on the valley splitting at filling factors $\nu = 3$ and $5$ (indicated by arrows), we plot in Fig. 5 (b) the $ln(R_{xx})$ of the SdH oscillation minima with temperature for $\nu = 3$ (blue circles) and $5$ (brown triangles). The linear dependence of $ln(R_{xx})$ on inverse temperature, represented by the fitted straight lines in Fig. 5 (b), confirms the thermally activated behaviour described by $R_{xx} \propto e^{\left(-\frac{\Delta_v}{2k_BT}\right)}$. From the slope of $ln(R_{xx})$ vs $1/T$, we extract the effective valley splitting $\Delta_v$. The inset of Fig. 5 (b) shows the extracted valley splitting which increases from $20\mu eV$ to $100\mu eV$ with the magnetic field which is consistent with the behaviour of the peak-to-peak spacing, $\Delta B$, described in section IV, Fig. 4 (b). In addition, we confirm the effective mass of an electron $m^* \approx (0.189 \pm 0.01)m_o$ in our system (see Supplementary Material SM-5), which is in agreement with the reported values [60].

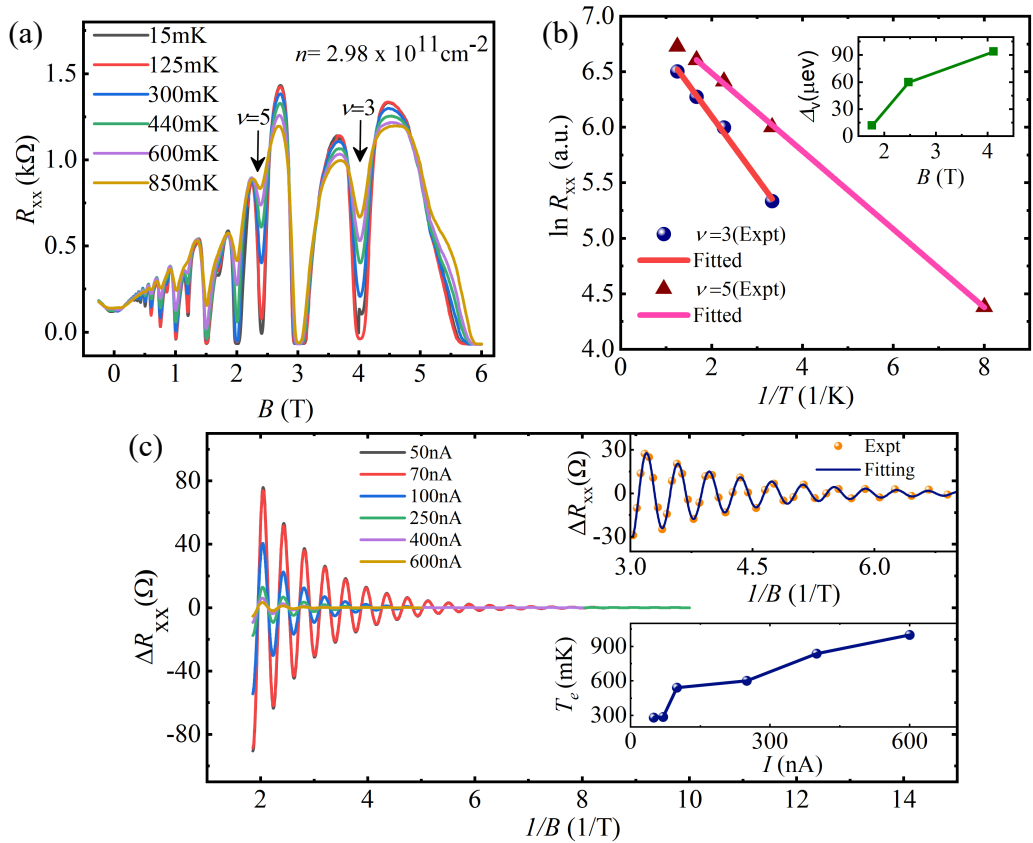

Fig. 5(a) Temperature dependence of SdH oscillation $(R_{xx})$ at $n = 2.98 \times 10^{11}$cm$^{-2}$ measured between 15 mK and 850 mK. The oscillation amplitude decreases with increasing temperature. (b) Temperature dependence of SdH oscillation minima plotted with $1/T$ at filling factors $\nu = 3,5$. Solid lines are the linear fits. Inset: The valley splitting extracted from the slope of the linear fits, increases with magnetic field from $\Delta_v \approx 20\mu eV$ to $100\mu eV$. (c) Heating effect of current biasing on SdH oscillations from 50nA to 600nA after background subtraction. The amplitude decreases with increasing current biasing. Top inset: The experimental data (scatter plot) shows agreement with the fitted expression $\Delta R_{xx} = -Aexp\left(-\frac{\alpha(T_b+\Delta T)}{B}\right)cos\left(\frac{2\pi F}{B}\right)$ (solid line) at $I = 50$nA. Bottom inset: The minimum carrier temperature extracted from the fitting saturates at $T_e \approx 300$mK.

We estimate the electron temperature on our device by inspecting the SdH peak broadening as a function of the driving current as shown in Fig. 5 (c). We observe a gradual suppression of the SdH oscillations, consistent with the thermal broadening, thereby confirming the current-bias induced heating. To extract the electron temperature, the oscillations are fitted to the semi-empirical expression [61,62]

$$\Delta R_{xx} = -A exp\left(-\frac{\alpha(T_b + \Delta T)}{B}\right) \cos\left(\frac{2\pi F}{B}\right)$$

where, $\alpha = \frac{2\pi^2 K_B m^*}{\hbar e}$ is the damping factor, $T_b$ is the bath temperature, $\Delta T \approx T_e - T_b$, and $m^*$ is the effective mass of the electron. The top inset to Fig. 5 (c) shows the experimentally observed SdH oscillations (scatter plot) and the solid line is a fit to the equation for $I = 50$nA, demonstrating a very good agreement with the experimental data. Bottom inset to Fig. 5 (c) shows the electron temperature, $T_e$, extracted from the fit. We find that $T_e$ decreases with the biasing current bias and plateaus at a minimum $T_e \approx 300$ mK. From the fit, the electron temperature at $I = 100$nA corresponds to a thermal broadening of $\approx 165\,\mu$eV, which is comparable to the spin-splitting energy at the onset field $B_s = 1$T discussed in section IV (Fig. 4a). This demonstrates that, although disorder-induced broadening decreases with increasing carrier density discussed in section III Fig. 3 (e), the thermal broadening limits the observation of both spin and valley splitting in the SdH oscillations.

## VI. CONCLUSIONS

In this manuscript we present a comprehensive analysis of the major scattering mechanisms and valley splitting energy in high-mobility undoped Si/SiGe 2DEG, which are critical indices for scalable spin-qubit implementations. The 2DEG maintains a carrier density in the range of $n \approx 10^{11}$cm$^{-2}$. The device exhibits a remarkably high mobility of $\mu = 1.6 \times 10^6$ cm$^2$V$^{-1}$s$^{-1}$ at $n = 7.09 \times 10^{11}$cm$^{-2}$ and remaining as high as $\sim 10^6$ cm$^2$V$^{-1}$s$^{-1}$ at lower carrier densities of $n \sim 3.1 \times 10^{11}$cm$^{-2}$ making these heterostructures promising for the realization of gated quantum dots with single-electron confinement. The power-law dependence of the mobility $\mu \propto n^\alpha$ with $\alpha \approx 0.6$ reveals that the remote impurity scattering is effectively screened by the 2DEG in the regime of interest, leaving background impurities near the quantum well as the remaining scattering source at high carrier concentration. Further comparison between transport and quantum lifetimes confirms that large angle scattering from background impurities near the quantum well dominates at a higher carrier concentration while small angle scattering due to the remote impurities are suppressed. Furthermore, we observe a sizable valley splitting in our devices, which is proportional to carrier concentration suggesting that the enhanced screening of impurity scattering by the 2DEG plays a major role in the observation of valley splitting. We also find an enhancement in the valley splitting with the magnetic field suggesting confinement effects also play a major role in their observation. The resolution of spin and valley splitting was found to be limited both by the disorder and thermally induced broadening. Through thermal activation measurements, we extract an the effective intrinsic valley splitting as large as $\sim 100$ $\mu$eV, surpassing the values reported in many unconfined lateral devices which is a key requirement for high-fidelity spin qubits.

## ACKNOWLEDGEMENTS

MT acknowledge funding support from the National Quantum Mission, an initiative of the Department of Science and Technology, Government of India, LLK acknowledges PMRF, MoE, Govt. of India for fellowship, KD, RZ and RHB thank the Deutsche Forschungsgemeinschaft (DFG) via the grant 469222030.

# Density-driven scattering and valley splitting in undoped Si/SiGe two-dimensional electron system

Lucky Donald Lyngdoh Kynshi[1], Umang Soni[1], Chithra H Sharma[2,3], Yu Cheng[4], Kristian Deneke[3], Robert Zierold[3], Shengqiang Zhou[4], Robert H Blick[3], Anil Shaji[1], and Madhu Thalakulam[1][*]

[1] Department of Physics, Indian Institute of Science Education and Research Thiruvananthapuram, Kerala 695551, India

[2] Institut für Experimentelle und Angewandte Physik, Christian-Albrechts-Universität zu Kiel, 24098 Kiel, Germany

[3] Center for Hybrid Nanostructures, Universität Hamburg, Luruper Chaussee 149, 22761 Hamburg, 22761 Germany

[4] Helmholtz-Zentrum Dresden-Rossendorf, Institute of Ion Beam Physics and Materials Research, Bautzner Landstraße 400, 01328 Dresden, Germany

## SM-1 Charge trapping induced hysteresis effect

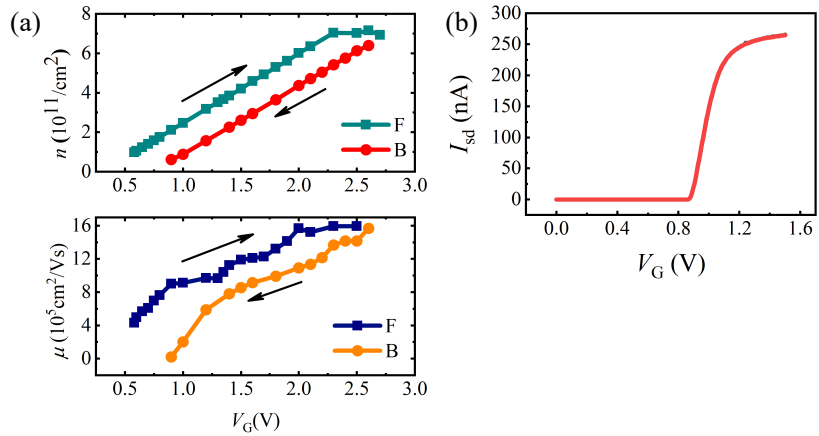

*Fig. SM-1. (a) Carrier concentration and mobility in the forward and backward directions. The black arrow indicates the direction of the gate voltage sweep. Both the carrier concentration and mobility exhibited hysteresis. (b) Subsequent turn-on characteristics changes to $V_{th} \approx$ 0.9V after the hysteresis effect.*

Figure SM-1(a) shows the carrier concentration and mobility as a function of the gate voltage measured during both forward (F) and backward (B) sweeps at 15mK. The black arrow indicates the direction of gate sweeping. Above $V_G = 2.3V$, both the carrier concentration and mobility saturate, which is attributed to charge trapping at the oxide-silicon interface. This trapping induces hysteresis, as evidenced by the backward sweep deviating from the forward trace. Figure SM-1(b) further illustrates that the subsequent gate sweeps exhibit a shift in the threshold voltage toward higher values of $V_G = V_{th} \approx 0.9V$. The magnitude of this shift depends on the maximum gate voltage applied in the saturation region.

[*] madhu@iisertvm.ac.in

# Remote and background scattering

Remote and background scattering are calculated using an equations described by Monroe et. al [1]

$$\mu_{remote} \cong \frac{16\sqrt{\pi g_s g_v}\, e\, n^{\frac{3}{2}}\, h_{eff}^3}{\hbar N_{remote}}$$

$$\mu_{background} \cong g_s^{\frac{3}{2}} g_v^{\frac{3}{2}} \frac{n^{\frac{1}{2}}}{4\pi\, \hbar\, N_{3D}}$$

$g_s = 2$ is the spin degeneracy

$g_v = 2$ is the valley degeneracy

$n$ is the carrier density

$h_{eff} \approx 50 nm$ is the height of the SiGe barrier

$e = 1.6 \times 10^{-19} C$ is the elementary charge

$N_{remote}$ is the remote average impurity density per unit area

$N_{background}$ is background impurity density per unit volume

In regime II, For carrier density $n \approx 1.05 \times 10^{11} cm^{-2}$, $N_{remote} \approx 1.22 \times 10^{12} cm^{-2}$

In regime III, For carrier density $n \approx 4.2 \times 10^{11} cm^{-2}$, $N_{background} \approx 1.06 \times 10^{13} cm^{-3}$

# SM-2 SdH oscillation background subtraction

Background subtraction of the Shubnikov–de Haas (SdH) oscillations was performed by fitting the oscillation extrema using a polynomial-fitting function [2]. Figure SM-2(a) shows the experimental raw longitudinal resistance $R_{xx}$ (grey curve), with the local maxima and minima fitted using polynomial functions (red and blue curves, respectively). The mean of these two fitted envelopes, represented by the green curve, defines a slowly varying background component of the $R_{xx}$ signal. The subtraction of this background from the raw $R_{xx}$ data yields the oscillatory component $\Delta R_{xx}$, as depicted in Fig. SM-2(b).

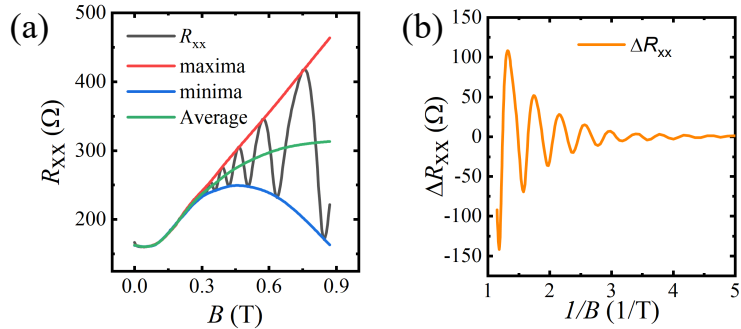

*Fig. SM-2. (a) Background subtraction method for the raw SdH oscillations (grey) fitted with envelope curves for maxima (red) and minima (blue). The green curve represents the average of both envelopes. (b) Background-subtracted SdH oscillation $\Delta R_{xx} = R_{xx} - R_b$ where $R_b$ is the average of the two fitted envelope functions.*

**SM-3 Transport and Quantum lifetime**

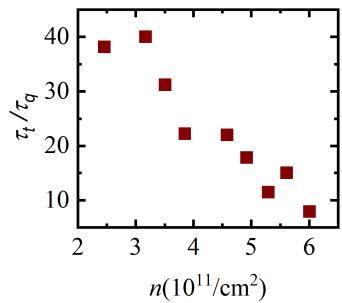

Fig. SM-3 Ratio of the transport lifetime '$\tau_t$' to the quantum lifetime '$\tau_q$'. The ratio $\frac{\tau_t}{\tau_q} > 1$ and it decreases with increasing carrier concentration.

The ratio of the transport lifetime to the quantum lifetimes is shown in Fig. SM-3(a). The ratio $\frac{\tau_t}{\tau_q} > 1$ and it decreases with increasing carrier concentration. This suggests that the 2DEG screens out all remote impurities, leaving only the background impurities near the quantum well at high carrier concentrations.

**SM-4 *g*-factor and valley splitting**

From the onset of the Landau and spin-resolved oscillations at magnetic fields $B_L$ and $B_s$ respectively, we extracted the effective g-factor by estimating the Landau level broadening ($\Gamma$), as $\Gamma \approx E_L$ at a magnetic field $B_L$, where $E_L$ is the energy spacing between the Landau levels . We use the relation

$$E_L(B_L) \approx E_z(B_s) \approx \Gamma \qquad (1)$$

$$\hbar\omega_c - g\mu_B B_L = \Gamma(B_L)$$
$$g\mu_B B_s = \Gamma(B_s) \qquad (2)$$

Incorporating the field dependent of the landau broadening [3–5], we have

$$\Gamma(B) \propto \sqrt{B}$$

$$\frac{\hbar\omega_c - g\mu_B B_L}{g\mu_B B_s} = \frac{\Gamma(B_L)}{\Gamma(B_s)}$$

Solving the equation, we have

$$g = \frac{2m_e}{m^*\left(1 + \sqrt{\frac{B_s}{B_L}}\right)} \tag{2}$$

Using equation (2), the effective g-factor with carrier concentration is shown in Fig. SM-4(a). The g-factor is higher than the bulk value of 2, which is attributed to the exchange interaction of electrons that enhances spin splitting. Moreover, the g-factor exhibited carrier concentration dependence, decreasing from 3.526 to 2.626 as the carrier concentration increased from $n = 2.45 \times 10^{11} \text{cm}^{-2}$ to $6 \times 10^{11} \text{cm}^{-2}$.

**Valley splitting at low carrier concentrations**

At low carrier concentrations from $n = 1.05 \times 10^{11} \text{cm}^{-2}$ to $n = 1.75 \times 10^{11} \text{cm}^{-2}$, as shown in Fig.SM-4(b), the valley splitting remain unresolved because of the significant Landau level broadening exceeding the valley splitting energy.

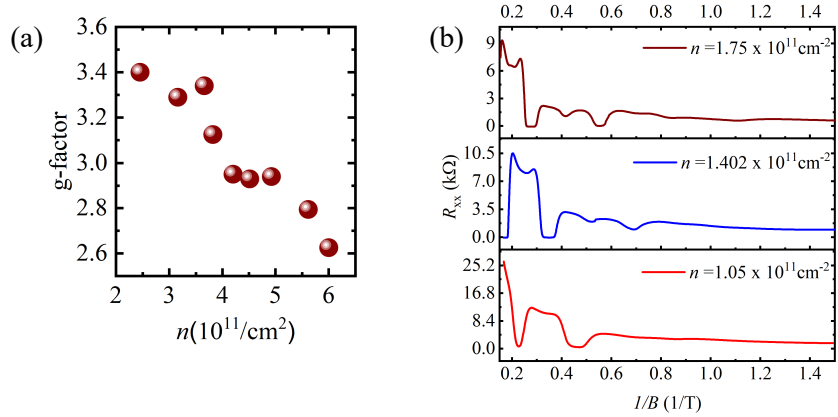

Fig. SM-4 (a) shows the effective g-factor extracted from the Landau and spin-resolved splitting. The g-factor decreases with increasing carrier concentration and approaches a bulk value of 2 at high carrier concentrations. (b)The oscillation with $1/B$ at lower carrier concentrations shows weak valley splitting because of the stronger Landau level broadening.

**SM-5 Effective mass**

Figure SM-5(a) shows the temperature dependence of the background-subtracted Shubnikov–de Haas (SdH) oscillations $\Delta R_{xx}$ measured in the range of 15 mK to 850 mK. The oscillation amplitude decreased with increasing temperature because of thermal broadening. We focus on two magnetic field values, $B=0.7$ T and $B = 0.565$ T, indicated by black arrows. At these fields, the temperature dependence of the oscillation amplitude was recorded and fitted using the equation [6,7]

$$\frac{\Delta R_{xx}(T)}{\Delta R_{xx}(T_o)} = \frac{T \sinh\left(\frac{2\pi^2 K_B T_o}{\hbar \omega_c}\right)}{T_o \sinh\left(\frac{2\pi^2 K_B T}{\hbar \omega_c}\right)}$$

where $\Delta R_{xx}(T)$ and $\Delta R_{xx}(T_o)$ are the amplitudes of the SdH oscillations at temperatures $T$ and $T_o$ ($T > T_o$) respectively. Here, $\omega_c = eB/m^*$ is the cyclotron frequency, $K_B$ is the Boltzmann constant, and $\hbar$ is the Planck's constant. Figure SM-5(b) shows the extracted amplitudes (scatter points) along with the fitted data (solid lines), yielding an effective mass of $m^* = (0.189 \pm 0.01)\, m_o$

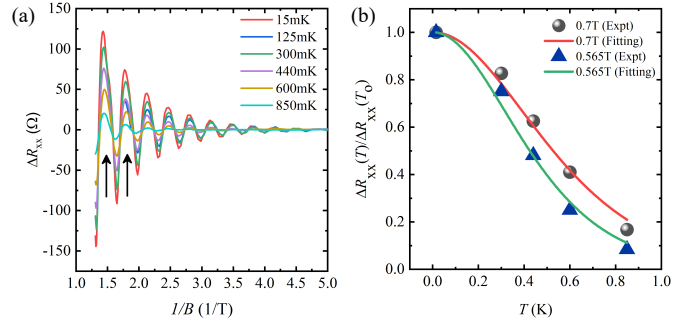

Fig. SM-5 (a) Temperature dependence of the background-subtracted Shubnikov–de Haas (SdH) oscillations $\Delta R_{xx}$ measured between 15 mK and 850 mK. (b) Extraction of the effective mass from the temperature-dependent SdH amplitude, showing experimental (points) and fitting (solid lines) data at two different magnetic field values of $B = 0.565T$ and $B = 0.7T$